\documentclass[12pt,aps]{revtex4}%
\usepackage{amssymb}
\usepackage{amsmath}
\usepackage{amsfonts}
\usepackage{graphicx}%
\providecommand{\U}[1]{\protect\rule{.1in}{.1in}}
\begin{document}
\title{Levi-Civita Cylinders with Fractional Angular Deficit}
\author{J.P. Krisch and E.N. Glass}
\affiliation{Department of Physics, University of Michigan, Ann Arbor, MI 48109}
\date{02 March 2011}

\begin{abstract}
The angular deficit factor in the Levi-Civita vacuum metric has been
parametrized using a Riemann-Liouville fractional integral. This introduces a
new parameter into the general relativistic cylinder description, the
fractional index $\alpha$. When the fractional index is continued into the
negative $\alpha$ region, new behavior is found in the Gott-Hiscock cylinder
and in an Israel shell.

\end{abstract}
\maketitle

\section{Introduction}

The\ 1917 Levi-Civita$^{\text{\cite{LC1917}}}$ solution is a standard vacuum
exterior for cylindrical matter distributions. The metric can be written in
terms of two parameters: mass density $\sigma\ $and angular deficit $b$
\begin{equation}
ds^{2}=-r^{4\sigma}dt^{2}+r^{-4\sigma}[r^{8\sigma^{2}}(dr^{2}+dz^{2}%
)+b^{2}r^{2}d\phi^{2}]. \label{L-C-met}%
\end{equation}
The interior matches add a mass/length parameter, $\lambda,$ to the
description of the complete solution.\ Currently 3-dim cylindrical matter
distributions with positive density are believed to provide a good description
for $0\leq\sigma<1/2$.$^{\text{\cite{Mar58}-\cite{BLSZ04}}}$\ The literature
has tended to focus on the $(\sigma,\lambda)$ ranges but one additional
feature in many of the interior solutions is the boundary location and size
that varies as a function of the density. The size of the circumference
remains finite even as the density increases and a coordinate radius blows
up.$^{\text{\cite{His85,Got85}}}$\ This suggests that the boundary
circumference is an interesting parameter to include in the analysis of
cylinder solutions.\ 

In this paper we consider a circumference calculated using a Riemann-Liouville
fractional integral.\ This introduces a new parameter into the cylinder
description, the fractional index $\alpha.$\textbf{\ }The circumference
calculated from the Levi-Civita metric involves the angular deficit factor. A
fractional circumference can relate the physical behavior of the angular
deficit to a mathematical framework through $b(\alpha)$. Matches to an
interior will provide a fractional parametrization of all of the cylinder
parameters.\ An unexpected result of the fractional extension is a new family
of solutions occuring at one of the angular deficit
embedding$^{\text{\cite{Got85}}}$ transition points.

In the next section we briefly review the literature leading to limits on the
range of $\sigma$, and develop the fractional circumference.$\ $The idea is
applied to the $\sigma=0$ Gott-Hiscock$^{\text{\cite{His85,Got85}}}$ constant
density string solution, providing a parametrization of $\lambda.$\ The method
is extended to a $\sigma\neq0$ shell$^{\text{\cite{WSS97}}}$ in the third part
of the paper and the behavior of $\lambda$ and $\sigma$ are linked. An
Appendix introducing some of the basic fractional integral definitions is
included. \ 

\section{Cylinder Solutions}

\subsection*{The range of $\sigma$}

The Riemann tensor of metric (\ref{L-C-met}) is zero for $\sigma=(0,$
$1/2)$\ and is singular as $r$ approaches zero%
\[%
\genfrac{}{}{0pt}{}{\lim}{r\rightarrow0}%
\left[  R_{abcd}R^{abcd}=\frac{64\sigma^{2}(4\sigma^{2}-2\sigma+1)(2\sigma
-1)^{2}}{r^{4(4\sigma^{2}-2\sigma+1)}}\right]  \rightarrow\infty.
\]
The literature contains a number of discussions that provide interior matter
distributions motivating some of this behavior.\ An infinite matter cylinder
is often considered because, in the small $\sigma$ limit, a test particle at
rest experiences an acceleration $\overset{..}{r\ }=-2\sigma/r$. This is the
Newtonian acceleration for a particle a distance $r$ from a line source of
mass/length$\ (\sigma)$.$^{\text{\cite{Phi96}}}$\ In addition, through metric
matching, $\sigma$ can often be related to the matter stress-energy coming
from the field equations.\ The identification is not absolute. The constant
density Gott-Hiscock solutions$^{\text{\cite{Got85,His85}}}$\ and the U(1)
string solutions of Garfinkle$^{\text{\cite{Gar85}}}$ both use the $\sigma=0$
Levi-Civita vacuum. The exact cylindrical\textbf{ }solid matter solutions that
can be matched to the Levi-Civita vacuum indicate a restricted range for
$\sigma,$ $0\leq\sigma<1/2.$ Some of the 3-dim matter solutions that can be
matched to $\sigma=0$ also show restricted ranges in the cylinder mass/length,
$\lambda,$ following from the field equations. Israel shells bounding vacuum
Levi-Civita and a second vacuum interior indicate a broader
range.$^{\text{\cite{WSS97}}}~$

One of the early investigations of static cylindrical solutions, due to
Marder$^{\text{\cite{Mar58}}}$, directly examined the relation between
$\sigma$ and the cylinder $\lambda$ by matching to a Levi-Civita vacuum with
no angular deficit.\ Krori and Paul$^{\text{\cite{KP77}}}$ used Marder's
solution to establish a limit $\sigma<1/2.$\ For general $\sigma,$
Davidson$^{\text{\cite{Dav90}}}$ developed a set of static cylindrical
solutions which were used by Bonnor et al$^{\text{\cite{BM91,BD92}}}$ to study
the possible range of $\sigma$. One of the motivating questions for this
discussion was the zero Riemann tensors for $\sigma=(0,$ $1/2)$. The
$\sigma=0$ flat spacetime was interpreted as a matter-free vacuum but the
$\sigma=1/2$ case was left an open question.\ Philbin$^{\text{\cite{Phi96}}}$
extended cylindrical solutions into the negative $\sigma$ region and suggested
that the endpoints $\left\vert \sigma\right\vert =1/2$ described planar rather
than cylindrical matter distributions. Based on the behavior of gyros orbiting
a cylindrical mass distribution, Herrera, Ruifern\'{a}ndez, and
Santos$^{\text{\cite{HRS01}}}$ also suggested that $\sigma=1/2$ described a
planar mass. In a later paper$^{\text{\cite{HST+01}}}$, Herrera et al treated
the $\sigma=1/2$ transition point in greater detail and, with the ($\varphi
,z$) coordinates taken as interchangeable to include possible planar
topologies, suggested models for the coordinate range\textbf{ }$0\leq
\sigma<\infty.$

Some of the work on static cylinders is embedded in discussions of rotating
cylinders with the parameter and density limits appearing for both cases. For
example, one of the early indications of a restricted relativistic density
range was the rotating dust cylinder of Vishveshwara and
Winicour.$^{\text{\cite{VW77}}}$ This interesting paper contains an expression
for $\lambda$ related to an angular deficit and identifies the $b=0$ deficit
factor as a critical limit related to a rotating column with $\lambda\leq1/4.$
\ Building on this work, Lathrop and Orsene$^{\text{\cite{LO80}}}$ considered
a cylinder with two counter-rotating dust currents and, for this matter
source, dupicated the Vishveshwara-Winicour$^{\text{\cite{VW77}}}$ density
limit, as did later work by da Silva et al.$^{\text{\cite{SH+95}}}$

\subsection*{Fractional Circumference}

A fractional integral is a function convolution over a range.\ The fractional
integral used to calculate the fractional circumference is the
Riemann-Liouville form$^{\text{\cite{OS74, MR93}}}$
\begin{equation}
I^{-\alpha}[f(x)]=\frac{1}{\Gamma(\alpha)}%
{\textstyle\int\limits_{0}^{x}}
f(y)(x-y)^{\alpha-1}dy
\end{equation}
where $\alpha$ is the fractional order, $Re(\alpha)>0$. A fractional
circumference is calculated by integrating $\sqrt{g_{\phi\phi}}$ around the
circle. $x$ is identified as $\phi$ with the circumference following in the
$\phi\rightarrow2\pi$ limit. For the base metric, consider a 3+1 Minkowski
metric with cylindrical coordinates $(t,r,\phi,z)$.%
\[
ds^{2}=-dt^{2}+dr^{2}+r^{2}d\phi^{2}+dz^{2}.
\]
The usual circumference of a circle in the $(r,\phi)$ plane for $r_{\ }%
=r_{0}\ $is $C=2\pi r_{0}.$ $r_{0}$ is the coordinate radius. A fractional
circumference for the same coordinate radius is%
\begin{equation}
C^{(\alpha)}=\frac{1}{\Gamma(\alpha)}%
\genfrac{}{}{0pt}{1}{\lim}{\phi\rightarrow2\pi}%
{\textstyle\int\limits_{0}^{\phi}}
\sqrt{g_{\phi\phi}}(\phi-y)^{\alpha-1}dy=2\pi r_{0}\frac{(2\pi)^{\alpha-1}%
}{\Gamma(1+\alpha)}.
\end{equation}
$\alpha=1$ gives the usual circumference. For $0<\alpha<1,$ the circumference
is less than the usual $2\pi r_{0}$ and for $\alpha>1$ it is larger. The
methods of fractional calculus$^{\text{\cite{Mai97,Pod99,BW00}}}$ have been
very successful in modeling transport processes with anomalous microscopic
time and/or spatial structure.$^{\text{\cite{SKB02, SGMR00}}}$ One could
expect that the density of matter distributions interior to the boundary would
reflect the transport processes responsible for their growth.\ In this case,
it is the variation in the circumference with $\alpha$ that we wish to link to
matter distributions with an angular deficit, with the fractional variation in
circumference reflecting a fractional surface matter distribution. If
coordinate ranges are not imposed, the angular deficit can be transformed
away. Here it is strongly linked to a physical description with the usual
imposed angular coordinate ranges. Bonner$^{\text{\cite{Bon79}}}$ noted that
the angular deficit parameter, $b$, determines the topology of the manifold
covered by metric (\ref{L-C-met}) and cannot be removed by scale
transformations. Angular deficit is a topological defect like the
gravitational Aharonov-Bohm effect discussed by Jensen and
Ku\v{c}era$^{\text{\cite{JK93}}}$

In the following, we apply the fractional circumference to the parameters of
the $\sigma=0$ Gott-Hiscock constant density string.\ 

\subsection*{The Gott-Hiscock String}

The Gott-Hiscock solution describes an interior constant density, $\delta,$
cylinder matched to a $\sigma=0$ Levi-Civita vacuum with angular deficit $b$.
The matter (-) and vacuum (+) metrics are%
\begin{align}
ds_{-}^{2}  &  =-c^{2}dt^{2}+d\rho^{2}+[\frac{\sin(\delta\rho)}{\delta}%
]^{2}d\phi^{2}+dz^{2}\label{matter-met}\\
ds_{+}^{2}  &  =-c^{2}dt^{2}+dr^{2}+b^{2}r^{2}d\phi^{2}+dz^{2} \label{vac-met}%
\end{align}
The string radii in the interior and exterior are $\rho_{0}$ and $r_{0}$.\ The
matching relations are $\sin(\delta\rho_{0})=\delta br_{0}$ , and
$b=\cos(\delta\rho_{0})$ and the mass per unit length, $\lambda,$ of the
constant density string is $\ \lambda=(1-b)/4.$ The angular deficit/excess
associated with $b$ is $\Delta\phi=2\pi(1-b)=8\pi\lambda.$ The allowed range
of $b$ is $-1\leq b\leq1$ and the positive $\lambda$ range is $0\leq
\lambda\leq1/2.$\ \ 

The boundary match provides an expression for the coordinate radius, $r_{0}$
and associated circumference \
\begin{align}
r_{0}  &  =\frac{\tan(\delta\rho_{0})}{\delta}\\
C  &  =2\pi r_{0}b=2\pi\frac{\sin(\delta\rho_{0})}{\delta}%
\end{align}
A $(r,\phi)$ cross section of the exterior vacuum is regarded as a circle with
a missing pie slice. $\ r_{0}b$ is the radius of the equivalent closed
circle.\ The matching relations indicate that $r_{0}$ approaches $\infty$ as
$\delta\rho_{0}\rightarrow\pi/2,$ while $b=0$ and the circumference takes its
maximum value.\ Gott$^{\text{\cite{Got85}}}$ has motivated this behavior by
embedding the $(t,z)=const$ Levi-Civita vacuum into a 3-dim metric with
coordinates $(w,r^{\prime}=br,\phi),$ the embedding relation between $w$ and
$r^{\prime}$ defining a cone in the 3-dim space.\ This embedding picture is
illustrated in the sketches of Figure 1. The shaded region represents a matter
cap in the interior. The sketches illustrate the entire $b$ range.\ In the
integer model, as the density increases from $\lambda=0$ $(b=1)$
to\ $\lambda=1/4$ the upper sides of the cone tilt outward to form a $b=0$
cylinder. As the mass continues to increase, the center moves out into the
vacuum, the lower sides tilting in to form an upward pointing
cone.\ Conventionally this is associated with negative $b$, angular excess and
a quasi-regular singularity$^{\text{\cite{VW77}}}$ in the vacuum.%

\begin{figure}[ptb]%
\centering
\includegraphics[
natheight=1.555800in,
natwidth=4.072600in,
height=1.5269in,
width=3.9508in
]%
{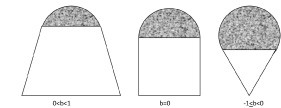}%
\caption{embedding diagrams for the GH string}%
\end{figure}

The circle defined by the matter/vacuum match has an effective radius $r_{0}b$
and is clearly maximum for the $b=0$ cylinder.

\subsubsection*{Fractional Angular Deficit}

The metric circumference of the string in the exterior spacetime is $2\pi
br_{0}$ where $br_{0}$ is the effective radius of the closed circle. Equating
this to the $\alpha>0$ fractional circumference one finds a fractional angular
deficit factor
\begin{align}
2\pi br_{0}  &  =2\pi r_{0}\frac{(2\pi)^{\alpha-1}}{\Gamma(1+\alpha)},\\
b(\alpha)  &  =\frac{(2\pi)^{\alpha-1}}{\Gamma(1+\alpha)}.\nonumber
\end{align}
$\alpha=1$ can be identified with $b=1$, the vacuum case.\ Fig.2 shows
$b(\alpha)\ $over the range $-0.5\leq\alpha\leq1$. $b$ decreases toward $0$ as
$\alpha\rightarrow0$ and $\lambda$ moves toward the cone $\rightarrow$
cylinder transition value, $\lambda=1/4.$%

\begin{figure}[ptb]%
\centering
\includegraphics[
natheight=4.407400in,
natwidth=4.588800in,
height=2.673in,
width=2.7822in
]%
{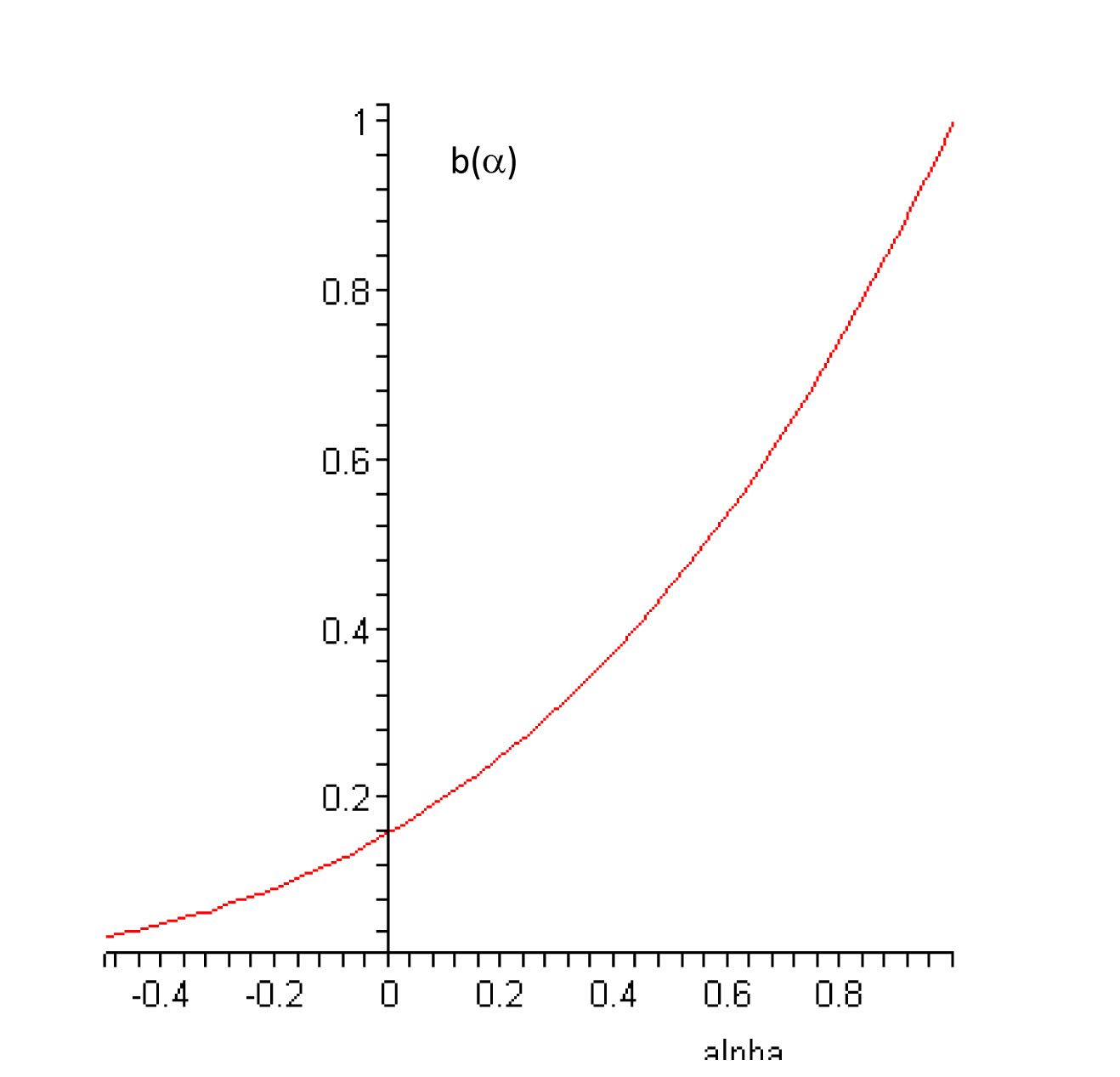}%
\caption{b vs $\alpha,$ $\alpha>-0.5$}%
\end{figure}

In this region, $br_{0}$, the effective radius in the circumference, increases
from its vacuum $0$ to its maximum value $br_{0}=1/\delta.$ This behavior is
implicit in the embedding diagrams. The value of $\alpha$ follows from
$\delta\rho_{0}=\sin^{-1}(br_{0}).$ \ In the $\alpha$ region shown in Fig.2,
the fractional index is only another parameter that provides a mathematical
base for $b$.\ However, while the fractional circumference is strictly defined
for $\alpha>0,$ the graph shows a continuation across the origin into the
undefined negative alpha region. The interpretation of negative $\alpha$ will
come from the string models it parametrizes.\ In Fig.2, there is no
identification of an $\alpha$ parametrization for $b=\ 0$ until the alpha
range is extended further along the negative axis.\ This extension is shown in
Fig. 3. There are a sequence of points of decreasing amplitude, oscillating
around $b=0.$ The fractional string model has only a small extension into the
negative $b$ region with its vacuum quasi-regular singularity.

The identification of a fractional form for $b$ can be extended to the string
mass/length showing $\lambda$ oscillations about the cone $\rightarrow$
cylinder critical value $\lambda=1/4.$

The mass/length is shown in Figures 4 and 5. Fig.4 supports the identification
of $b=1$, $\alpha=1$ with a zero string vacuum and Fig.5 shows the expected
oscillations in the string mass/length about $\lambda=1/4.$ The string
coordinate radius, $r_{0}=\tan(\delta\rho_{0})/\delta,$ blows up at
$\delta\rho_{0}=\pi/2$ \ but the product $br_{0}$ entering into the
circumference is finite, showing small oscillations around $r_{0}b=1/\delta$,
the effective radius for the cylindrical embedding. This example used a 3-dim
matter distribution inside the cylinder.\ The next example considers a
cylindrical shell.%
\begin{figure}[ptb]%
\centering
\includegraphics[
natheight=4.407400in,
natwidth=4.588800in,
height=2.673in,
width=2.7822in
]%
{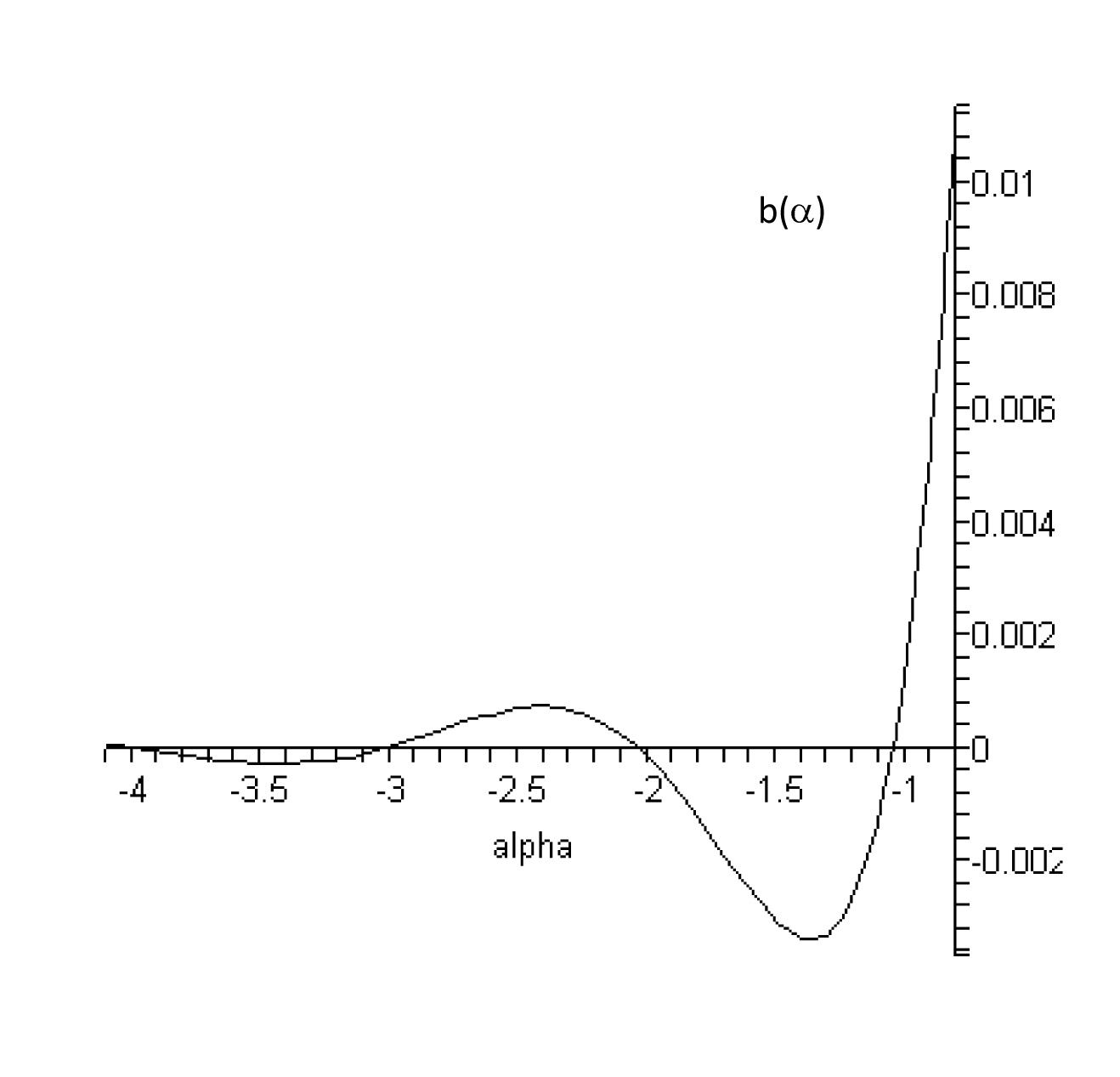}%
\caption{b vs $\alpha,$ $-0.5<\alpha$}%
\end{figure}

%

\begin{figure}[ptb]%
\centering
\includegraphics[
natheight=4.463400in,
natwidth=4.829700in,
height=2.673in,
width=2.8904in
]%
{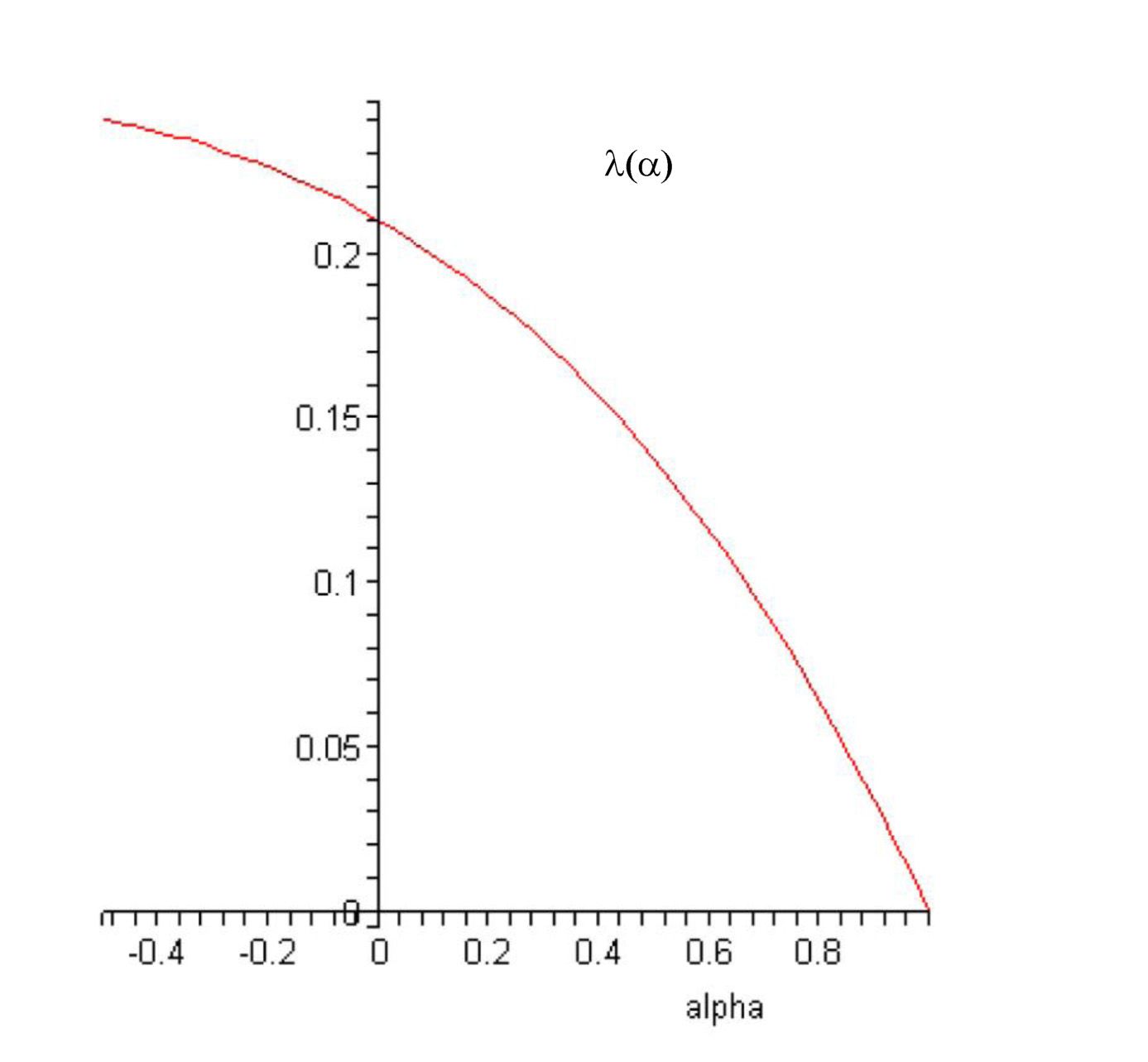}%
\caption{$\lambda$ vs $\alpha,$ $-0.5<\alpha$}%
\end{figure}
%

\begin{figure}[ptb]%
\centering
\includegraphics[
natheight=4.459700in,
natwidth=4.872100in,
height=2.673in,
width=2.9175in
]%
{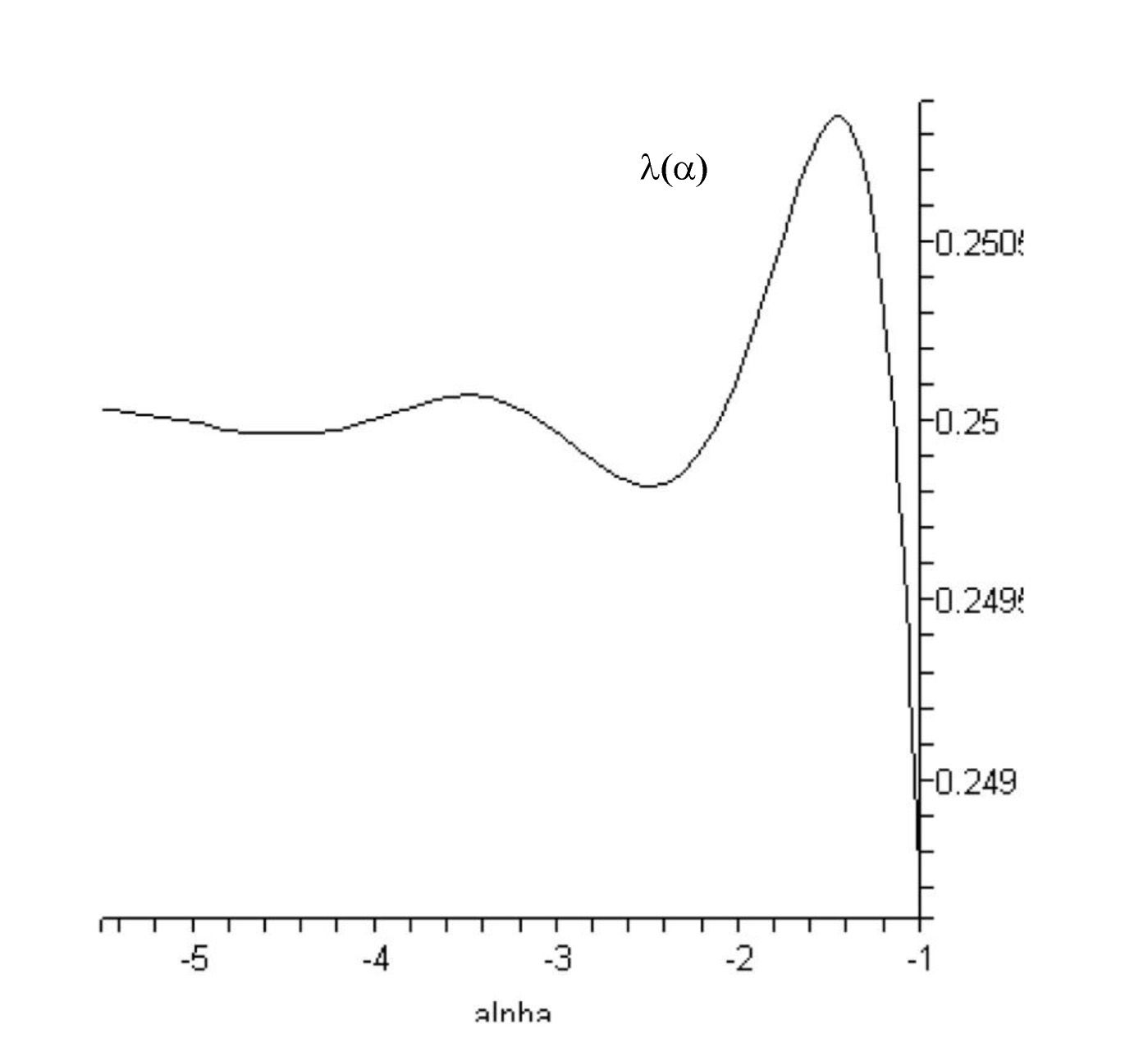}%
\caption{$\lambda$ vs $\alpha,$ $\alpha<-0.75$}%
\end{figure}

\section{Cylindrical Shells}

2+1 cylindrical shells have also been used to explore the static
parameter/density limits. \ A sequence of cylindrical shells was first
suggested by Marder$^{\text{\cite{Mar58}}}$ and Wang et
al$^{\text{\cite{WSS97}}}$ discussed a shell with stress-energy resulting from
a jump from exterior Levi-Civita to an interior Minkowski. A simple
cylindrical model with non zero $\sigma$ has exterior\ (+) and interior (-)
metrics
\begin{align}
ds_{+}^{2}  &  =-(r/r_{0})^{4\sigma}dt^{2}+(r/r_{0})^{-4\sigma}[(r/r_{0}%
)^{8\sigma^{2}}(dr^{2}+dz^{2})+b^{2}r^{2}d\phi^{2}]\\
ds_{-}^{2}  &  =-dt^{2}+d\rho^{2}+dz^{2}+\rho^{2}d\phi^{2}%
\end{align}
The metric match across a surface $(r_{0},\rho_{0})$ is%
\begin{equation}
\rho_{0}=br_{0}%
\end{equation}
This matching relation suggests that the circumference is bounded as $b$ and
$r_{0}$ change. The extrinsic curvatures are $K_{ab}^{\pm}=n_{a;b}^{\pm
}=-\Gamma_{ab}^{^{\pm}r}n_{r}^{\pm}=\partial_{r}g_{ab}/2$ at $r=r_{0}.$\ The
shell has stress-energy, $S_{ab}$, related to the extrinsic curvature jump
$<K_{ab}>\ =K_{ab}^{+}-K_{ab}^{-},$ $<K>\ =\ <K_{a}^{a}>$, across the
boundary.$^{\text{\cite{Isr66, Poi04}}}$\ Calculating the stress-energy one
finds, with $\digamma(\sigma)=1-4\sigma+4\sigma^{2}$
\begin{align}
-8\pi S_{j}^{i}  &  =\ <K_{j}^{i}>-h_{j}^{i}<K>\\
-8\pi S_{t}^{t}  &  =\frac{1-b\digamma(\sigma)}{br_{0}}\\
-8\pi S_{\phi}^{\phi}  &  =-\ 4\sigma^{2}/r_{0}\\
-8\pi S_{z}^{z}  &  =\frac{1-b}{br_{0}}%
\end{align}
Using the same simple density integral to calculate $\lambda$ as was used in
the Gott-Hiscock string, one finds the stress-energy parameters%
\begin{align}
4\lambda &  =1-b\digamma(\sigma)\\
8\pi P_{\phi}  &  =4\sigma^{2}/r_{0}\\
8\pi P_{z}  &  =\frac{b-1}{br_{0}}\\
8\pi\varepsilon &  =\frac{1-b\digamma(\sigma)}{br_{0}}%
\end{align}
The general equation of state$^{\text{\cite{Cle94,BLSZ04}}}$ is
\begin{equation}
\varepsilon+P_{\phi}+P_{z}=4\sigma/r_{0}%
\end{equation}
Looking first at $\sigma=0,$ the usual stiff string equation of state is
found.\ This is an interesting comparison case to the GH string.\ It does
describe a constant density object with an axial tension and $\lambda$ but the
limits are set by the angular deficit factor, not by trignometric
limits.\ Using ad hoc values $b=(1,0,-1)$, we have $4\lambda=(0,1,2)$. Thus we
have the same set of boundary values as the GH string.\ With $b=(2\pi
)^{\alpha-1}/\Gamma(1+\alpha)$ the mass/length has the same form as for the GH
string
\begin{equation}
4\lambda=1-\frac{(2\pi)^{\alpha-1}}{\Gamma(1+\alpha)}%
\end{equation}
and the same density oscillations around $b=0$ are present. \ A large
difference between the two cases is that $b$ is not limited to a unit range,
without the GH identification of $b$. The current limits on $\sigma$ would
restrict $b$. For $\sigma\neq0$ the immediate observations are, that for fixed
radius, $\sigma$ determines the tangential stress while the axial stress is
determined by $b$.\ Links between $\sigma$ and $b$ follow from assumptions on
the partial equations of state linking $\varepsilon$ and individual stresses,
or on relations between the stresses.\ For example, a shell with uniform
stress, $P_{\phi}=P_{z\text{ }}$imposes $b=1/(1-4\sigma^{2})$. \ In this
model, isotropic stress is associated with negative mass/length.\ A physical
solution will have anisotropic stresses, avoiding a singularity at
$\sigma=1/2$.

\section{Conclusion}

In this paper we have examined a fractional parametrization $b(\alpha)$ of the
angular deficit factor.\ The original motivation was to match the physical
behavior of the angular deficit to a mathematical framework providing a
fractional parametrization of cylinder parameters.\ For the Gott-Hiscock
cylinder, we found new behavior when the angular deficit factor is continued
from a vacuum with a conical defect across the embedding cylinder, $b=0,$ into
a vacuum with a quasi-regular singularity. The behavior may have implications
for the stability of the cylinder solutions in the $b$ parameter
range.$^{\text{\cite{HL94,PAM10}}}$ This is under investigation.\ The models
considered in the paper are for the Levi-Civita $\sigma$ in the range
$0\leq\sigma<1/2.$ The applicability of a fractional angular deficit as
developed here, depends on an angular coordinate with an associated
circumference.\ Herrera et al$^{\text{\cite{HST+01}}}$ used coordinates
$(\varphi,z)$ with ranges $-\infty$ to $\infty,$ imposing a periodicity,
$\varphi=\varphi+2\pi$ on the angular coordinate for $\sigma<1/2$. Our result
applies in that region.\ Their results seem to imply that $\sigma=1/2$ is
associated with a planar topology and that there is no appropriate angular
coordinate.\ For $\sigma>1/2$ they suggest that the Levi-Civita coordinates
($\varphi,z$) are interchanged, with $z$ becoming the periodic measure,
$z=z+z_{0}$. The circumference for this $\sigma$ region would be
\[
C^{(\alpha)}=\frac{1}{\Gamma(\alpha)}%
\genfrac{}{}{0pt}{1}{\lim}{z\rightarrow z_{0}}%
{\textstyle\int\limits_{0}^{z}}
\sqrt{g_{zz}}(z-y)^{\alpha-1}dy=\frac{(z_{0})^{\alpha}}{\Gamma(1+\alpha)}%
\]
It is also possible that both $z$ and $\varphi$ are periodic in this region
suggesting a toroidal matter distribution rather than a cylindrical one.\ This
extended view would allow two fractional circumferences to be included in the
modeling. The complete topological description of Levi-Civita matched matter
distributions, and its fractional extensions, still have many open
questions\textbf{.} The infinite cylinder, both static and rotating, continues
to be an interesting and useful structure in general relativity: \ The
parameter effects due to including a cosmological constant in
Levi-Civita$^{\text{\cite{Lin86,Tia86}}}$ are beginning to be
examined$^{\text{\cite{GS10}}}$ and, with the recent connections to braneworld
cosmologies$^{\text{\cite{Sak09,HS10}}}$, relativistic cylinders have acquired
new physical relevance. \ 

\appendix{}

\section{FRACTIONAL CALCULUS}

Fractional calculus was invented in 1695 when
L'Hopital$^{\text{\cite{OS74,MR93}}}$ asked Leibniz about the meaning of his
notation $\frac{d^{n}y}{dx^{n}}$ for $n=1/2$. \ This question, also asked by
Bernoulli defined the idea of a fractional derivative and\ L'Hopital's
question about derivatives has developed into a fractional calculus, a
framework for integrals and derivatives of non-integer order. \ Euler made the
first real contribution to the development of fractional calculus with his
1738 proof that the fractional derivative of $x^{n}$ was
meaningful.$^{\text{\cite{SKM93}}}$ The first contribution using the idea is
credited to Abel for his 1823 work on the tautochrone.$^{\text{\cite{MR93}}}%
$\ The idea of fractional calculus has attracted the attention of many of the
same people that are associated with the development of physics and general
relativity: Lagrange, Laplace, Fourier, Liouville, Riemann and
Weyl$^{\text{\cite{OS74,MR93}}}$ among others. There are several definitions
of fractional integrals depending on the integral limits. The Riemann form
integrates from $c$ to $x$, The Liouville version from $-\infty$ to $x$, and
in this paper the Riemann-Liouville form is used from $0$ to $x$%
.$^{\text{\cite{MR93}}}$ \
\begin{equation}
I^{-\alpha}[f(x)]=\frac{1}{\Gamma(\alpha)}%
{\textstyle\int\limits_{0}^{x}}
f(y)(x-y)^{\alpha-1}dy
\end{equation}
$\alpha$ is the fractional order, $Re(\alpha)>0.$\ There is also a Weyl form
for the fractional integral with a $+\infty$ limit. \ 

In order for the Riemann-Liouville integral to converge it is necessary that
$f(x)$ satisfy the condition
\begin{equation}
f(x^{-1})=O(x^{1-\delta})\text{ \ }\delta>0
\end{equation}
Functions obeying this condition are Riemann class; for example, the function
$x^{b},\ b>-1$ is a Riemann class function. The Weyl integral is useful for
functions with the convergence property%
\begin{equation}
f(-x)=O(x^{-\nu-\delta})\text{ }\delta>0,\ x\rightarrow\infty
\end{equation}
Functions satisfying this condition are said to be of Liouville class; for
example, $x^{-b}$ with $b>\nu>0,$ is a Liouville function. The allowable
parameter range eliminates constants from this class. There can be overlap
between classes. \ 

The evaluation of fractional integrals uses the beta and gamma functions. The
gamma function is defined by%
\begin{align}
\Gamma(x)  &  :=%
{\displaystyle\int\limits_{0}^{\infty}}
e^{-t}t^{\alpha-1}dt,\text{ \ }x>0\\
\Gamma(x+1)  &  =x\Gamma(x),\text{ \ }x>0\nonumber
\end{align}
For positive integer $n$, $\Gamma(n+1)=n!$. $\Gamma(x)$ is extended to
negative $x$ by the functional equations%
\begin{align}
\Gamma(x)  &  :=\frac{\Gamma(x+1)}{x},\text{ \ }-1<x<0\\
\Gamma(x)  &  :=\frac{\Gamma(x+2)}{x(x+1)},\text{ \ }-2<x<-1\nonumber\\
\Gamma(x)  &  :=\frac{\Gamma(x+n)}{x(x+1)\cdot\cdot\cdot\text{ }%
(x+n-1)},\text{ \ }-n<x<-n+1\nonumber
\end{align}

The beta function is defined as
\begin{equation}
B(z,w)=\int\limits_{0}^{1}x^{z-1}(1-x)^{w-1}dx
\end{equation}
It is related to the gamma function
\begin{equation}
B(z,w)=\frac{\Gamma(z)\Gamma(w)}{\Gamma(z+w)}%
\end{equation}
For example, the fractional integral of $f(x)=1$ is%
\begin{equation}
D_{0}^{-\nu}[1]=\frac{1}{\Gamma(\nu)}\int\limits_{0}^{x}(x-y)^{(\nu-1)}dy
\end{equation}
Comparing with the definitions of the beta and gamma functions, we see the
fractional integral of $1$ is given by
\begin{equation}
D_{0}^{-\nu}[1]=\frac{\Gamma(1)}{\Gamma(\nu+1)}x^{\nu}\
\end{equation}

Another example is the fractional integral of $f(x)=x^{\alpha}.$\ Using the
definitions the integral becomes%
\begin{equation}
D_{0}^{-\nu}[x^{\alpha}]=\frac{\Gamma(1+\alpha)}{\Gamma(\alpha+\nu+1)}%
x^{\nu+\alpha}\
\end{equation}
For $\nu=1,$ this becomes $x^{\alpha+1}/(\alpha+1)$, the usual integral of
$x^{\alpha}.$ For $\alpha=-1,$ the gamma function in the numerator is singular
so the fractional integral of $1/x$ is not defined. The analytic continuation
used in the paper assigned meaning to the singular points of the gamma function.

\end{document}